\documentclass{iopart}
\input{epsf}
\usepackage{graphicx}

\begin{document}

\title{Destruction of Anderson localization by nonlinearity
in kicked rotator at different effective dimensions} 

\author{L. Ermann$^1$ and D.L. Shepelyansky$^2$}
\address{$^1$ Departamento de F\'isica, GIyA, 
Comisi\'on Nacional de Energ\'ia At\'omica, Buenos Aires, Argentina}
\address{$^2$Laboratoire de Physique Th\'eorique du CNRS, IRSAMC, 
Universit\'e de Toulouse, UPS, 31062 Toulouse, France}

% \eads{\mailto{$^1$ermann@irsamc.ups-tlse.fr}, \mailto{$^2$dima@irsamc.ups-tlse.fr}}

%\date{\today}
%\date{Received  11 March 2014}

\begin{abstract}
We study numerically the frequency modulated kicked nonlinear rotator
with effective dimension $d=1,2,3,4$. We follow 
the time  evolution of the model up to $10^9$ kicks
and determine the exponent $\alpha$ of subdiffusive spreading
which changes from $0.35$ to $0.5$ when the dimension
changes from $d=1$ to $4$. All results are obtained in a regime
of relatively strong Anderson localization
well below the Anderson transition point 
existing for $d=3,4$. We explain
that this variation of the exponent is 
different from the usual $d-$dimensional Anderson models 
with local nonlinearity where $\alpha$ drops with increasing $d$.
We also argue that the renormalization arguments proposed by
{\it Cherroret N et al. arXiv:1401.1038} are not valid.

\end{abstract}

%Quantum chaos; semiclassical methods
% }
%Fiber lasers, 42.55.Wd
%Optical fibers, 42.81.-i
%Thermodynamics, 05.70.-a
%Anderson localization disordered solids, 71.23.An
%statistical physics and nonlinear dynamics, 05.10.-a
%05.30.Ch 	Quantum ensemble theory 
%05.45.-a 	Nonlinear dynamics and chaos 
%05.70.Ce 	Thermodynamic functions and equations of state

\vskip 0.3cm
Dated: March 11, 2014

\submitto{\JPA}
\maketitle

\section{Introduction}

At present there is a significant interest
to effects of nonlinearity on Anderson localization 
\cite{anderson}. The early theoretical and numerical
studies \cite{dls1993,molina} have been followed by
more recent and more detailed analysis performed by different groups
\cite{danse,flach2009prl,flach2009kg,garciapre,flach2010pre,flach2010epl},
\cite{bodyfelt,aubry2010,marioepl2010,fishmanpre2011,marionjp2013,flach2013}.
The interest to this problem comes also from the side of mathematics
which puts forward a fundamental question on how
the pure point spectrum of Anderson
localization is affected by a weak nonlinearity 
\cite{wang,bourgain,fishman2012}. At the same time the experiments on
spreading of light in nonlinear photonic lattices \cite{segev,silberberg}
and of Bose-Einstein condensates of cold atoms 
in disordered potential \cite{inguscio}
start to be able to observe effects of nonlinearity on localization.

The main effect found in numerical simulations is a subdiffusive spreading
of wave packet over lattice sites induced by a moderate nonlinearity.
Large time scale simulations are required to determine the spreading exponent 
with a good accuracy and hence the choice of a good model, which is
easy for numerical simulations and at the same time captures 
the  main physical effects, is important. 
One of such models is the model of kicked nonlinear rotator \cite{dls1993},
where nonlinear phase shifts are introduced in 
the quantum Chirikov standard map, known also as the kicked rotator
\cite{chirikov1981}. 

It is also important 
that the kicked rotator has one more interesting
extension: the frequency modulated kicked rotator (FMKR)
introduced in \cite{dls1983}. In this model the kick amplitudes are modulated
with $d-1$ incommensurate frequencies that
allows to model the Anderson transition in effective dimensions
$d=3,4$ \cite{casati,borgonovi,dls2011}.
This FMKR model, proposed theoretically, 
has been realized 
in skillful and impressive experiments 
with cold atoms  by Garreau group \cite{garreau1}. 
These experiments allowed to observe the Anderson transition 
in $d=3$ and to determine experimentally the critical exponents
which have been found to be in agreement with analytical 
and numerical calculations \cite{garreau2}.
At the moment the Garreau experiments are definitely
represent the most advanced  experimental 
investigation of the Anderson transition
both in fields of cold atoms and solid state disordered systems.

In a recent preprint \cite{garreau3} it is proposed to
to study frequency modulated kicked nonlinear rotator
(FMKNR) model. It is argued there that the FMKNR allows to
investigate the effects of nonlinearity of 
Anderson transition in $d=3$. Here, we show that
the renormalization group analysis performed in \cite{garreau3}
is not relevant for the main physical effects 
leading to the nonlinearity induced wave spreading.
However, the investigation the FMKNR model itself is
interesting and provides some new information
on effects of nonlinearity on Anderson localization.
Thus we present here the results of our numerical studies of
FMKNR in effective dimensions $d=1,2,3,4$ 
up to times $t=10^9$. The model is described in Sec.2,
numerical results are presented in Sec.3, simple estimates 
are presented in Sec.4 and discussion is given
in Sec.5.

\section{FMKNR model description}

The time evolution of the wave function of the FMKNR
is described by the equation
\begin{equation}
 \psi(t+1)=e^{-i(\hat{H}_{0}+\hat{H}_{nl})}e^{-i\hat{V}(t)} \psi(t) \; ,
\label{eq1}
\end{equation}
where 
\begin{equation}
  \hat{H}_{0}(n) = \xi_n \;\; (model \; M1) \; ; \;\;\; 
   \hat{H}_{0}(n) = Tn(n+\zeta)/2  \;\; (model \; M2) \; ,
\label{eq2}
\end{equation}
with $\xi_n$ being random energies distributed
homogeneously in the interval $-\pi, \pi$ (model $M1$)
and $H_0(n)=Tn(n+\zeta)/2$ (model $M2$) 
are rotational phases in a kicked rotator
with $\zeta$ corresponding to a quasi-momentum of Bloch waves in kicked
optical lattices and parameter $T \sim \hbar_{eff} \sim1$.
Here, $n$ is a quantum number corresponding to momentum quantization
\cite{dls1983,casati,borgonovi,dls2011,garreau1,garreau2}.
This part of Hamiltonian describes a free propagation.

\begin{figure} 
\begin{indented}\item[]
\begin{center}
\includegraphics[width=0.82\textwidth]{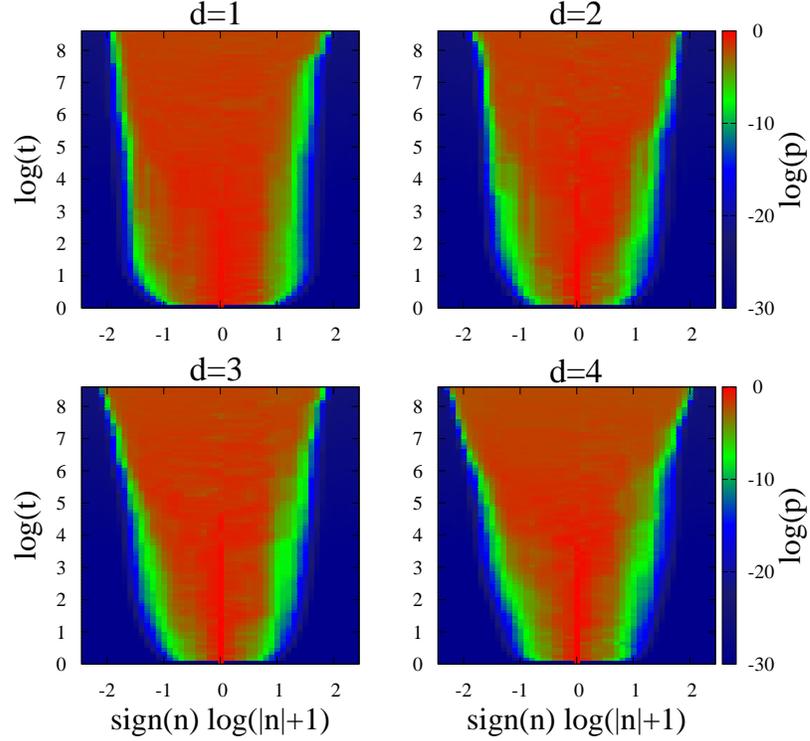}
\end{center}
\vglue 0.1cm
\caption{(Color online) 
Time evolution of the  probability spreading $p=|\psi(n)|^2$, shown by color,
over momentum levels $n$  for the model $M1$ at $k=1.5$ ($d=1$);
$k=0.5, \epsilon=0.75$ ($d=2,3,4$), and $\beta=1$, the time interval
is $1 \leq t \leq 10^9$. Initial state is $n=0$.
Probability $p$ is shown by color variation given 
by color bars. The logarithms are decimal.
}
\label{fig1}
\end{indented}
\end{figure} 

The nonlinear phase shift, as in \cite{dls1993},  is given by
\begin{equation}
  \hat{H}_{nl} = \beta |\psi(n)|^2 \; ,
\label{eq3}
\end{equation}
where $\beta$ is the strength on nonlinear interactions
and $\psi(n)$ is taken in the momentum representation.
The norm $\sum_n |\psi(n)|^2=1$ is conserved by unitary evolution.

The part with the kick is written in the phase representation
for $\psi(\theta)$ which is conjugated to the momentum representation
($\psi(\theta+2\pi)=\psi(\theta)$, ${\hat n} = - i \partial /\partial \theta$):
\begin{equation}
  V(\theta, t) = k \cos \theta [1+\epsilon \Pi_{i=1}^{d-1} \cos (\omega_i t)] \; .
\label{eq4}
\end{equation}
Here, $\epsilon$ represents a strength of frequency modulation
with $d-1$ incommensurate frequencies $\omega_i$ and $t$ is measured in a number of 
map iterations. For $d=1$ in $M2$ case
we obtain the usual kicked rotator \cite{chirikov1981} with $k \sim 1/\hbar_{eff}$,
$T \sim \hbar_{eff}$ and the chaos parameter $K=kT = const$
($\hbar_{eff}$ is an effective Planck constant).
For $d=2$ we use $\omega_1 = 2\pi/\lambda$, for
$d=3$ we use $\omega_1  = 2\pi/\lambda$, $\omega_2=2\pi/\lambda^2$.
For $d=4$ we add frequency $\omega_3=2\pi/\sqrt{2}$.
Here $\lambda=1.32471795724475$ is a root of cubic equation \cite{borgonovi}.
At $\beta=0$ the models $M1$ and $M2$ manifest 
the phenomenon of  Anderson localization in 
effective dimensions $d=1,2,3,4$; for $d=3,4$
there is the Anderson transition for $k>k_c$ at a certain fixed $\epsilon$
\cite {borgonovi,dls2011,garreau1,garreau2}. For $d=3$ the curve
of the Anderson transition in the plane $(k,\epsilon)$ is
analyzed in \cite{delande}.

For $\beta \sim 1$ the model $M1$ at $d=1$ (KNR) shows a subdiffusive spreading
of wave packet over sites (levels) with the second moment growth
characterized by an exponent $\alpha \sim 0.4$ \cite{dls1993}:
\begin{equation}
  \sigma = \sum_n |\psi(n)|^2 n^2 =<n^2> \sim t^\alpha.
\label{eq5}
\end{equation}
It was argued \cite{dls1993} that this model effectively describes 
the spreading in discrete Anderson nonlinear Schr\"odinger equation (DANSE).
The later studies indeed confirmed that in DANSE the exponent
$\alpha$ is approximately the same as in the KNR \cite{danse,garciapre,flach2010epl}. The examples of probability spreading in the FMKNR at $d=1,2,3,4$
in the model $M1$ are shown in Fig.~\ref{fig1}.

The study of the FMKNR at $d=3$ has been proposed in \cite{garreau3}.
At $\beta=0$ the model FMKR can be exactly mapped on an Anderson model
in effective dimension $d$ \cite{casati,borgonovi}
by a transformation similar to those used
in $d=1$ case \cite{prange,dls1987}.
Indeed, since the phases $\theta_i=\omega_i t$ rotates with fixed frequencies
we can write the Hamiltonian in effective extended dimension $d$:
\begin{equation}
  H(n,n_i, \theta, \theta_i) = H_0(n) + \sum_{i=1}^{d-1} \omega_i n_i +
  \beta \sum_{n_i=-\infty}^{\infty} |\psi(n, n_i)|^2 + V(\theta, \theta_i) \delta_1(t)
\label{eq6}
\end{equation}
where $\delta_1(t)$ is a  periodic delta-function with period unity,
$n,\theta$ and $n_i, \theta_i (i=1,..,d-1)$ are conjugated pairs of variables.
Then the evolution is given by the unitary propagator:
\begin{equation}
  \psi(n,n_i, t+1) = 
\exp(-i H_{int}(n, n_i))\exp(-i V(\theta, \theta_i)) \psi(n,n_i,t) \; ,
\label{eq7}
\end{equation}
where $ H_{int}(n, n_i) =H_0(n)+ \sum_{i=1}^{d-1} \omega_i n_i +
  \beta \sum_{n_{i}=-\infty}^{\infty} |\psi(n, n_i)|^2$.
It is important here that the nonlinear term with $\beta$
contains a sum over all additional effective dimensions $d-1$.
In a certain sense this corresponds to long-range interaction
of planes in $d-1$ dimensions. This corresponds to the physics of the FMKNR model
where nonlinear coupling acts only in $n$ independently of
phases $\theta_i=\omega_{i}t$.

If we would model a real nonlinear interaction term in dimension $d$
we would have another Hamiltonian
\begin{equation}
  H(n,n_i, \theta, \theta_i) = H_0(n) + \sum_{i=1}^{d-1} \omega_i n_i +
  \beta  |\psi(n, n_i)|^2 + V(\theta, \theta_i) \delta_1(t)
\label{eq8}
\end{equation}
where nonlinear term $ \beta  |\psi(n, n_i)|^2$ have no summation
over $n_i$. Then the evolution of $\psi$ is still given by the 
propagator (\ref{eq7}) but with 
$ H_{int}(n, n_i) =H_0(n)+ \sum_{i=1}^{d-1} \omega_i n_i 
+  \beta  |\psi(n, n_i)|^2$.
Such a local term appears in DANSE in $d=2$
and has been studied in \cite{garciapre}. The numerical results
of \cite{garciapre} show that the exponent $\alpha$ 
of the second moment $n^2 \propto t^\alpha$ decreases
when we increase $d$ from $d=1$ to $d=2$ going from $\alpha \sim 0.4$ down to 
$\alpha \sim 0.25$. The analytical arguments of  \cite{garciapre}
give:
\begin{equation}
  \alpha = 2/(3d+2) \; .
\label{eq9}
\end{equation}
Of course, this expression assumes local nonlinear interaction term
as in (\ref{eq8}) that is rather different from the case
of long interactions in effective dimensions effectively appearing in
the FMKNR of (\ref{eq1}), (\ref{eq6}). 

All renormalization group arguments presented in \cite{garreau3}
are developed for the case of local nonlinear term (\ref{eq8})
while the numerical simulations are done for the 
FMKNR case (\ref{eq1}) and (\ref{eq6}) corresponding to 
the long-range interactions.
Due to such a mixing of concepts the arguments of \cite{garreau3}
are not valid. Also, we point our that the renormalization group arguments
\cite{garreau3} assume a proximity to a critical point 
of the Anderson transition. But all the studies of the nonlinearity
induced destruction of Anderson localization show that 
it takes place even in a relatively strong localization regime 
and also in $d=1,2$ where the Anderson transition is absent
and linear waves are always localized.
Due to that reasons we argue that the approach of \cite{garreau3}
is not valid for the physics of phenomenon of DANSE.
However, the investigation of the FMKNR model at certain $d$ 
is interesting and thus we present our results below for $d=1,2,3,4$.

\section{Numerical results}

In our numerical studies we fix $k=1.5$ for $d=1$
and $k=0.5$, $\epsilon=0.75$  for $d=2,3,4$
for both models $M1$ and $M2$
The frequencies $\omega_i$ are fixed at
values given in the previous Section.
For the model $M2$ we use $T=2.89$ as in experiments
\cite{garreau3}; we use up to $10$ random values of quasi-momentum $\zeta$
in model $M2$ $(0 \leq \zeta  <1$) 
and up to $10$ disorder realizations in model $M1$.
The initial state is taken at $n=0$.
The transition between momentum and angle representations is done
by the fast Fourier transform.

\begin{figure} 
\begin{indented}\item[]
\begin{center}
\includegraphics[width=0.82\textwidth]{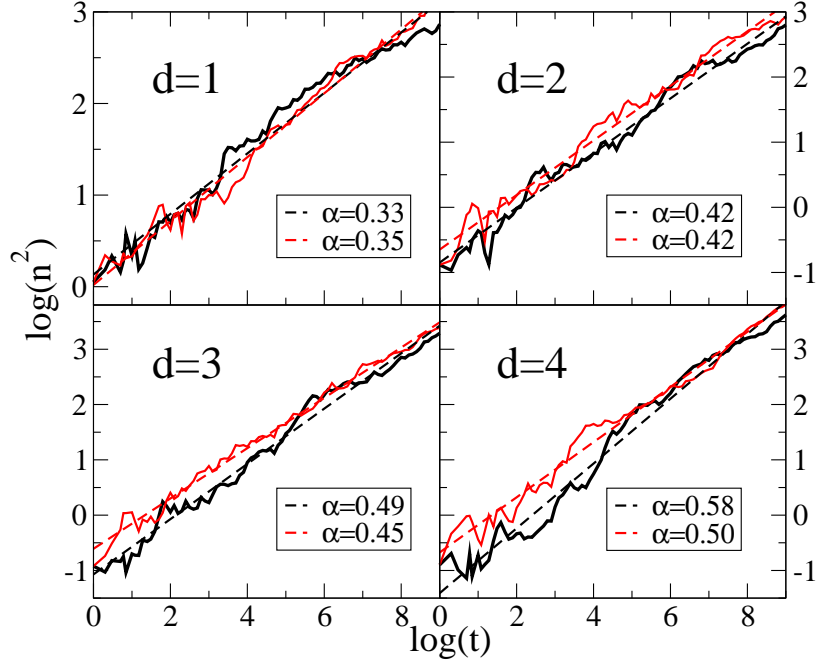}
\end{center}
\vglue -0.0cm
\caption{(Color online) Dependence of the second moment 
$\sigma\equiv\langle n^2\rangle$
of probability distribution  on time $t$ shown in 
logarithmic scale for  models $M1$ (black) and $M2$ (red) 
with dimensions $d=1,2,3,4$. Parameters $k, \epsilon, \beta$ 
are as in Fig.~\ref{fig1} and $T=2.89$ for $M2$. 
The power law fit of
subdifussive spreading  $\sigma\sim t^\alpha$ 
is shown by the straight dashed lines for each model. 
Effective dimensions $d$ and fitted values of $\alpha$ are shown 
in each panel, logarithms are decimal.
}
\label{fig2}
\end{indented}
\end{figure} 

For $d=3,4$ we find that both models have approximately 
the same critical value $k_c$ of Anderson transition.
For $d=3$ we have $k_c \approx 1.8$ 
in agreement with \cite{borgonovi,dls2011}.
Also for $d=4$ both models have
the same critical point $k_c \approx 1.15$ at $\epsilon = 0.9$
and $k_c \approx 1.3$ at $\epsilon =0.75$.
At $d-1$ frequencies the classical chaos border becomes very small
in $k$ so that random rotational phases in model $M1$
have the Anderson transition approximately at the same point as
in the model $M2$. We stress that all amplitudes of $k$ 
used in our simulations are located in a well localized phase
being rather far from the Anderson transition in $d=3,4$.
At that $k$ values the localization length
$\ell \sim 1 - 2$ captures only a couple of sites
(see Figs.1,9 in \cite{borgonovi}).

The spearing of probability $p_n=|\psi(n)|^2$
over momentum modes $n$ is shown as a function of time
$t \leq 10^{9}$ in Fig.~\ref{fig1} for model $M1$. The data show that
$|\psi(n)|^2$ spreads more or less homogeneously over 
a plateau (chapeau) which width increases with time.

The growth of the second moment 
$\sigma$ with time $t \leq 10^9$ is shown in Fig.~\ref{fig2}
for models $M1$ and $M2$ for a one disorder realization in $M1$
and one value of quasi-momentum  in model $M2$
(a random value in the interval $0\leq \zeta <1$).

\begin{figure} 
\begin{indented}\item[]
\begin{center}
\vglue 1.8cm
\includegraphics[width=0.82\textwidth]{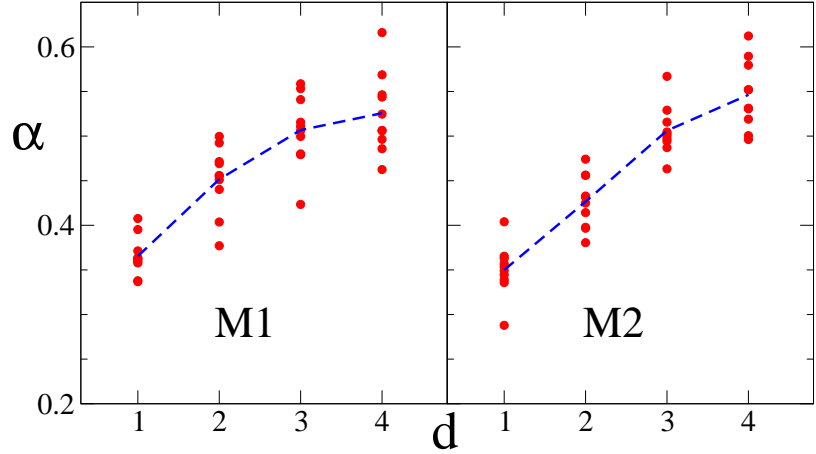}
\end{center}
\vglue 0.0cm
\caption{(Color online) The subdifussive spreading 
exponent $\alpha$ for dimensions $d=1,2,3,4$ of both models.
The model parameters are as in Figs.~\ref{fig1},~\ref{fig2}.
The exponents $\alpha$ are computed up to time $t=10^8$
for $10$ random realizations
of disorder in model $M1$  and $10$ random values of 
quasi-momentum $\zeta$ in model $M2$
(left and right panels respectively).
Dashed blue curves represent average $\alpha$ dependence on 
dimension $d$ for each model.
}
\label{fig3}
\end{indented}
\end{figure} 

We also determine the dependence $\sigma(t)$ for $1 \leq t \leq 10^8$ for 
both models for $10$ realizations of disorder ($M1$) or
$10$ random values of $\zeta$ ($M2$). The values of exponent $\alpha$
are shown in Fig.~\ref{fig3} for all four dimensions $d$.
Averaging over these 10 values of $\alpha$ we find the average
$\alpha$ value and its error-bar. 
For model $M1$ we obtain 
$\alpha=0.36\pm0.02$, $0.45\pm0.03$, $0.50\pm0.04$, $0.52\pm0.04$
 and for model $M2$
we obtain $\alpha=0.35\pm0.03$, $0.43\pm0.03$, $0.51\pm0.03$, $0.54\pm0.04$
respectively for $d=1,2,3,4$. Within the error-bars both models have the same 
value of $\alpha$ for a given dimension.

We note that the case $d=3$ for model $M2$ has been studied
in \cite{garreau3} with numerically obtained value
$\alpha \approx 0.4$. However, the time scales
considered there are about 1000 times smaller than those considered here.
Also in \cite{garreau3} the working point was placed rather 
close to the Anderson transition so that the localization length
of linear problem was rather large so that it was more difficult to reach
the asymptotic regime (in our case we are far from
the Anderson transition point and $\ell \sim 1 - 2$).

Our data show a clear tendency of growth of $\alpha$
with $d$ in the FMKNR model (\ref{eq1}). Of course, this
dependence is absolutely different from the 
one of (\ref{eq9}) obtained for a local nonlinear term.

\section{Simple estimates}

It is interesting to note that the exponent $\alpha=1/2$
corresponds to a so-called regime of ``strong chaos'' 
\cite{flach2010epl,bodyfelt,basko}. Indeed, the numerical simulations
performed in \cite{flach2010epl,bodyfelt} introduced 
a randomization of phases of linear eigenmodes after a fixed time scale
$\tau \sim 1$ showing numerically that in such a case $\alpha=1/2$.
This relation can be understood on a basis of simple 
estimates in a following way: the equations of amplitudes
of linear modes $C_m$ in the interaction representation have a form 
$i \partial C/\partial t \sim \beta C^3$ \cite{dls1993,danse,garciapre}.
In \cite{dls1993,danse,garciapre} it was assumed that
there is a plateau in amplitudes of $C \sim 1/(\Delta n )^{1/2}$ 
with $|m| < \Delta n$ and $C=0$ outside of the plateau.
Then the time scale $t_s$ after which a next level outside of plateau
will be populated is estimated as 
$\Gamma \sim 1/t_s \sim \beta^2 C^6 \sim \beta^2 /(\Delta n)^3$
due to norm conservation. Since the diffusion coefficient is
$D \sim (\Delta n)^2/t \sim \Gamma$ this gives
$\alpha =2/5$ for ($d=1$) and the relation (\ref{eq9}) 
for any $d$ \cite{dls1993,garciapre}.

It is also possible to assume that there is a certain
smooth profile distribution of $C$ values on the plateau
and use the estimate of the Fermi golden rule type \cite{landau}
used in quantum mechanics with 
$\Gamma \sim \beta^2 C^4 \sim \beta^2/(\Delta n)^2$
that would lead to $\alpha=1/2$ and $\sigma \sim t^{1/2}$
in agreement with arguments of \cite{flach2010epl,bodyfelt,basko}.
This assumes random phase approximation and 
mixing of phases on a certain fixed time scale $\tau$.
Thus in such a case we can write
 \begin{equation}
   (\Delta n)^2 \sim (t/\tau)^{1/2} \; . 
\label{eq10}
\end{equation}
However, it is clear that the time scale should grow with $\Delta n$
since the rate of chaotization should become smaller and smaller
with time since nonlinear term decreases.
The most natural assumption is that $\tau \sim 1/\delta \omega \sim
\Delta n$ where $\delta \omega \sim \beta |\psi_n|^2$
is a nonlinear frequency shift.Thus using the relation
$\tau \sim \Delta n$ we obtain from (\ref{eq10}) that again
$\alpha=2/5$.

The numerically obtained values of $\alpha$ (see Fig.~\ref{fig3})
are approximately located in the range $0.35 \leq \alpha \leq 0.5$.
It is possible that for $d=3,4$ a larger number of
modulation phases generates a more dense spectrum 
which is more similar to random phase approximation with $\tau \sim const$
corresponding to the strong chaos regime with $\alpha=1/2$. 
It is also possible that times even longer than $t=10^9$
are required to be in a really asymptotic regime.

\section{Discussion}

We present the results of numerical studies of the FMKNR models
with nonlinearity
in effective dimensions $d=1,2,3,4$. Our results show that
the exponent $\alpha$ of subdiffusive spreading
increases from $\alpha \approx 0.35$ up to $\alpha \approx 0.5$
when $d$ changes from $1$ to $4$. We show that this dependence on $d$
corresponds to a regime of nonlinearity with a long-range
interactions typical for FMKNR. In contrast to FMKNR, for 
Anderson models, with local nonlinearity like for DANSE \cite{danse,garciapre},
we have a decrease of the exponent $\alpha$
with increase of $d$  given by the relation (\ref{eq9}).

In our opinion, the exact derivation of the expression for the exponent
$\alpha$ represents  a nontrivial problem, Indeed, the results
presented in \cite{fishmanpre2011} clearly show that the
measure of chaos decreases with a growing system size.
This important result leads us to  a conclusion
that a spreading proceeds over more and more tiny
chaotic layers of smaller and smaller measure.
In such a regime a role of correlations should be important
and exact derivation of the expression for $\alpha$
requires additional information about
a structure of chaotic layers in many-body 
nonlinear systems.

\end{document}